# Tunable Hyperbolic Metamaterials Based on Self-Assembled Carbon Nanotubes

*John Andris Roberts[1]‡, Shang-Jie Yu[2]‡, Po-Hsun Ho[3], Stefan Schoeche[4], Abram L. Falk[3]\*, and Jonathan A. Fan[2]\**

[1]Department of Applied Physics, Stanford University, Stanford, CA 94305, U.S.A.

[2]Department of Electrical Engineering, Stanford University, Stanford, CA 94305, U.S.A.

[3]IBM T.J. Watson Research Center, Yorktown Heights, NY 10598, U.S.A.

[4]J.A. Woollam Co., Inc., Lincoln, NE 68508, U.S.A.

‡ These authors contributed equally to this work.

\*To whom correspondence should be addressed (alfalk@us.ibm.com, jonfan@stanford.edu)

**Abstract**

We show that packed, horizontally aligned films of single-walled carbon nanotubes are hyperbolic metamaterials with ultra-subwavelength unit cells and dynamic tunability. Using Mueller-matrix ellipsometry, we characterize the films' doping-level dependent optical properties and find a broadband hyperbolic region tunable in the mid-infrared. To characterize the dispersion of in-plane hyperbolic plasmon modes, we etch the nanotube films into nanoribbons with differing widths and orientations relative to the nanotube axis, and we observe that the hyperbolic modes support strong light localization. Agreement between the experiments and theoretical models using the ellipsometry data indicates that the packed carbon nanotubes support bulk anisotropic responses at the nanoscale. Self-assembled films of carbon nanotubes are well suited for applications in thermal emission and photodetection, and they serve as model systems for studying light-matter interactions in the deep subwavelength regime.

**Main Text**

Hyperbolic metamaterials (HMMs) are optically anisotropic media that have permittivities of opposite sign along different coordinate axes.[1-3] They are capable of localizing light to extreme subwavelength dimensions and can support a large optical density of states[4] across a wide spectral range. In addition, they can serve as model systems for studying extreme light-matter interactions, such as photonic topological transitions.[5] HMMs are actively being investigated for a number of application areas, including super-resolution imaging,[6] broadband absorption,[7] thermal emission control,[8] optical sensing,[9, 10] and ultra-fast optical sources.[11]

An ideal HMM platform would simultaneously support ultra-subwavelength unit cells, tailorable properties via material customization, and dynamic tunability. However, achieving these characteristics together has proven to be challenging. HMMs that consist of metallic and dielectric nanostructure assemblies, such as stacks of metal-dielectric thin films[5, 8, 9, 11] and aligned metallic nanowires in a dielectric matrix,[10, 12] are not dynamically tunable and possess relatively large unit cell dimensions. Naturally hyperbolic materials,[13, 14] such as hexagonal boron nitride (h-BN),[15-17] α-phase molybdenum trioxide,[18, 19] and bismuth selenide, exhibit hyperbolic behavior down to atomic-scale dimensions and can possess extremely low optical absorption. However, these materials have fixed hyperbolic spectral ranges, and many require materials growth or fabrication processing that are difficult to scale to macroscopic areas.

In this study, we present packed, self-assembled, and horizontally aligned single-walled carbon nanotube (SWCNT) films as a new class of dynamically tunable HMM. SWCNTs have several key advantages that make them excellent building blocks for designer HMMs. First, individual SWCNTs have distinct physical properties that depend on their diameter and chirality.[20] Composite films with customizable optical properties can therefore be produced by mixing together and assembling different SWCNT types.[21] Second, SWCNTs have single-nm diameters, and aligned SWCNT films can therefore be treated as a true effective medium down to extremely subwavelength size scales.[22] Third, semiconducting SWCNTs have optical properties that can be tuned through various gating or doping methods, allowing SWCNT HMMs to be dynamically tuned. Fourth, horizontally aligned SWCNTs can support in-plane hyperbolic propagation, which is difficult to achieve in stacked-layer or nanowire-based HMMs and has been the subject of recent studies.[19, 23] Fifth, SWCNTs have high thermal conductivities and are tolerant to very high temperatures, making them suitable as thermal emitters. We note parallel and complementary work on the thermal emission properties of SWCNT-based hyperbolic nanostructures at high temperatures.[24, 25]

Our study conceptually builds on prior theoretical studies of hyperbolic materials based on nanotube forests[26-28] and uses a recently developed self-assembly technique that utilizes controlled vacuum filtration.[21, 24, 29-33] This method has recently yielded films that are monolithic and globally aligned across the wafer scale,[29] electrostatically tunable,[33] and hexagonally-ordered in cross section when the packing fraction is high.[34] Highly ordered and large area SWCNT films have proven to be a unique platform to study nanoscale light-matter interactions, including intersubband plasmons,[33] hybrid phonon-plasmon modes,[30, 31] ultrastrong exciton-cavity coupling,[32, 34] and polarization-sensitive mid-infrared (MIR) and terahertz emission and detection.[10, 29, 30] In this study, we focus specifically on the anisotropic bulk optical properties of these films. Fundamentally, these properties emerge from the optical properties of the individual nanotubes[35] and the degree to which charges can be transported across nanotubes, which is an important open question in nanotube plasmonics.

To probe the hyperbolic nature of our SWCNT system, we investigate both the bulk optical properties of the aligned SWCNT metamaterial, using ellipsometry, and the behavior of hyperbolic plasmon modes (HPMs) confined in nanoribbon resonators patterned from the metamaterial. The confined HPMs have characteristics that derive from propagating HPMs in a continuous film. Since the aligned SWCNT film thicknesses are much smaller than the mid-infrared free space wavelengths representing the hyperbolic spectral range, these HPMs are strongly confined in the nanotube film and decay into the substrate and air cladding (Figure 1a, left). Because the alignment axis of the SWCNT film is parallel to the substrate, the HPMs exhibit in-plane hyperbolic dispersion. HPMs propagating at an angle relative to the nanotube alignment axis can have very large modal wavevectors $\beta$ (Figure 1a, right). As a result, the HPMs can confine light to deeply subwavelength scales.

The patterning of the SWCNT films into nanoribbons enables the direct coupling of free space radiation into confined HPMs and the study of their in-plane hyperbolic dispersion characteristics. The nanoribbons function as Fabry-Pérot (FP) resonators that confine propagating HPMs (Figure 1b). These plasmons propagate across the nanoribbon resonators and reflect at the ribbon edges, and the lowest order FP resonance results when these reflection events constructively interfere and the condition $2\beta L + 2\phi = 2\pi$ is satisfied. The wavevector $\beta$ is always perpendicular to the nanoribbon cut direction, and the term $2\beta L$ is the propagation phase accumulated from a round trip across the nanoribbon. The $2\phi$ phase term derives from plasmon reflection at the ribbon edge. At resonance, $\beta = (\pi - \phi)/L$ is specified by fixing the nanoribbon width. By measuring the resonant frequency of the nanoribbons as a function of $L$ and cut angle relative to the SWCNT alignment axis, we can map out the dispersion of the HPMs.[30, 36] This FP resonator model is equivalent to a treatment of the resonators as a low-order two-dimensional hyperbolic cavity.[4, 15, 37]

To experimentally prepare our SWCNT HMM system, we utilize a vacuum-filtration technique with SWCNTs that are approximately 1/3 metallic and 2/3 semiconducting, and we transfer the resulting films to a silicon substrate with a thin hafnia overlayer. The metallic tubes enable a plasmonic response in the bulk film, even without an external doping source.[30] The average length of the SWCNTs is approximately 500 nm. The as-assembled film has a 1-inch diameter (Figure 2a, inset). Scanning electron microscopy (SEM) (Figure 2a), cross-sectional transmission electron microscopy (TEM) (Figure 2b), and x-ray diffraction[34] show that the nanotubes are highly aligned and, in cross section, hexagonally ordered, with a 1.74 nm inter-tube spacing. The local hexagonal ordering of these packed films is confirmed by Fourier transforming the TEM image (Figure 2b, inset), which shows clear peaks corresponding to a hexagonal lattice arrangement.[34] Together, the images of the aligned SWCNTs at different length scales reveal a truly multi-scalar hyperbolic system.

We utilize Mueller matrix spectroscopic ellipsometry to characterize the dielectric function tensor of bulk SWCNT films. Under the assumption of uniaxial anisotropy with optical axis alignment along the length of the nanotubes, we apply a parameterized oscillator model to extract the hyperbolic dielectric function tensor components (see Supporting Information for more details).[38, 39] We consider films with two different doping levels: a "low doping-level" film with a resistivity of $4 \times 10^{-3}$ ohm-cm, which is measured in ambient conditions without intentional doping, and a "moderate doping-level" film, which is doped through exposure to $HNO_3$ vapor and has a resistivity of $1.6 \times 10^{-3}$ ohm-cm (see Supporting Information).[30]

These ellipsometry measurements reveal that our SWCNT films are type-I HMMs,[3] with a negative permittivity along the nanotube-alignment axis and positive permittivities along the other axes, across a wide range of infrared wavelengths (Figure 2c). In the low doping-level film, the spectral ranges of the hyperbolic dielectric tensor are 0.09 – 0.3 eV (4 – 14 μm) and 0.7 – 0.8 eV (1.6 – 1.8 μm). The negative permittivity in the low energy range (< 0.3 eV) can be understood as the response of free charge carriers from the metallic SWCNTs and low-doped semiconducting SWCNTs. The Lorentzian peaks around 0.6 eV and 1.2 eV correspond to the $S_{11}$ and $S_{22}$ exciton bands, respectively, in the semiconducting SWCNTs.[40] In between, there is a relatively narrow 0.7-0.8 eV negative permittivity region originating from the first interband plasmon resonance of the semiconducting SWCNTs.[41] In the moderately doped sample, the exciton absorption is suppressed due to band filling deriving from the increased free charge carrier density.[42] Here, the hyperbolic spectral range is 0.2 – 0.8 eV (1.7 – 6.7 μm), a dramatic change from that of the low-doped film.

To pattern these films into nanoribbon resonators and directly probe the HPM modes of the SWCNT system, we use electron-beam lithography and reactive ion oxygen etching, with fabrication parameters taken from ref. [30]. A series of nanoribbon arrays, each with a 200 μm × 200 μm area, are fabricated from an individual 95nm thick SWCNT film (Figure 3a), with $L$ spanning 55 nm to 395 nm and cut angles $\theta$ (defined in Figure 3b) ranging from 90° (cut normal to nanotube axis) to 20°. The SEM images (Figure 3b) display ribbons with uniform dimensions and well-defined edges, indicating the high quality of our fabrication process flow and our ability to pattern nanoscale features in our SWCNT films with high fidelity.

To characterize the HPM ribbon resonators, we employ micro-Fourier transform infrared (μ-FTIR) spectroscopy on the arrays, with the incident light polarization parallel to the SWCNT axis. The high degree of thickness uniformity (plus/minus 10%) and homogeneity across the 1-inch SWCNT film make it possible to measure different arrays of the same film with consistent results. We measure the extinction spectra 1-$T/T_0$ of the ribbon resonator arrays, where $T$ is the transmittance of the sample on the substrate and $T_0$ is the transmittance of the bare substrate. Each resonator array is measured at different gas doping levels (high, moderate, and annealed), which correspond to resistivities of $8 \times 10^{-4}$, $1.6 \times 10^{-3}$, and $8 \times 10^{-3}$ ohm-cm, respectively (see Supporting Information).

Our measurements of HPM nanoribbon resonators show that when ribbons of fixed width are cut at angles further from normal to the SWCNT axis ($\theta < 90°$), the plasmon resonances significantly redshift. This redshift indicates high mode confinement (i.e., large HPM wavevector $\beta$). While one might attribute this redshifting simply to the lengthening of the SWCNTs for $\theta < 90°$, this effect is actually a result of in-plane hyperbolic dispersion, as we show below. For moderately-doped ribbons with $L = 230$ nm, we observe that as the cut angle decreases from 90° to 20°, the resonance redshifts from ~0.5 eV to ~0.1 eV (Figure 3c). These resonators can support extremely subwavelength resonances, and ribbons with $L = 55$ nm are resonant at wavelengths as large as $\lambda/L \approx 70$ for $\theta = 20°$, compared to $\lambda/L \approx 38$ for $\theta = 90°$.

Theoretical spectra obtained using finite difference time domain (FDTD) and finite element method (FEM) simulations show excellent agreement with the experimental results (Figure 3c-d). In this analysis, the SWCNTs are treated as an anisotropic medium with a uniaxial dielectric tensor extracted from ellipsometry (Figure 2c). The agreement between the simulated and experimental spectra indicates that the aligned SWCNT film functions as a metamaterial with a bulk effective medium response. The simulated field profiles show that, although the incident polarization is along the nanotube axis, the excited mode has electric field components both perpendicular and parallel to the nanotubes. This characteristic indicates the hyperbolic nature of the mode (Figure S8). The ultrafine spatial granularity of the nanotube film is particularly important when examining the field profiles of the nanoribbon plasmon modes. Figure 3e shows a map of the electrical field strength |**E**| in a $\theta = 90°$ nanoribbon on resonance (0.5 eV). These field plots reveal strongly varying electric fields with length scales of only tens of nanometers. Thus, the single-nm unit cell dimension set by the SWCNT diameter is key to the films' performance as an effective medium. In contrast, HMMs with highly heterogeneous structuring at these length scales would be strongly affected by non-local effects.[43]

We obtain experimental dispersion relations of the HPMs by recording the peak resonant frequencies of nanoribbons with differing cut angles and widths and calculating the wavevector for each nanoribbon as $\beta = (\pi - \phi)/L$ (Figure 4a). We set a constant HPM reflection phase term, $\phi = (0.28 \pm 0.07\pi)$, based on calculated values (Figure S9).[44] As a comparison, we theoretically model the dispersion of the HPMs in the SWCNT film using the transfer-matrix method[45, 46] and the moderate doping dielectric functions measured by ellipsometry (Figure 4b). The theoretical and experimental dispersion curves show good agreement.

The large change in the dispersion relations as a function of doping demonstrates the excellent tunability of the SWCNT HMM (Figure 4c). For a resonator array with $L = 95$ nm and $\theta = 60°$, the resonance blueshifts approximately an octave, from $\omega \approx 2480$ cm$^{-1}$ to 5270 cm$^{-1}$, as the doping level increases. Such tunability suggests that photonic topological transitions[5] and epsilon-near-zero modes[47] could be dynamically controlled in these carbon nanotube systems across the MIR-wavelength range. While we employ chemical doping using HNO$_3$ vapor, alternative methods based on electrical gate tuning could also be used,[33] with conventional electronic gates for sufficiently thin HMMs and ion gating for thicker, bulk HMMs.

To further visualize the hyperbolic behavior of our carbon nanotube system, we extract and plot the experimental in-plane isofrequency contour in wavevector space at $\omega = 4000$ cm$^{-1}$. To create this contour, we identify nanoribbons of differing widths and cut angles that are approximately resonant at our target frequency (dashed horizontal line, Figure 4a). We then plot the in-plane wavevector components of each nanoribbon, defined as $\beta_z = \beta\sin\theta$ and $\beta_x = \beta\cos\theta$, where the z-axis is along the nanotubes (Figure 1a). The isofrequency contour is plotted in Figure 4d and reveals a clear hyperbolic shape. As $\theta$ decreases, higher-$\beta$ points on the isofrequency contour are sampled and $L$ decreases accordingly. At $\theta \approx 40°$, the hyperbolic curve asymptotes and the largest wavevector is achieved. For comparison, we also theoretically model the isofrequency contour using the transfer-matrix method and the ellipsometry data for moderate doping (see Supporting Information). The theoretical isofrequency contour shows good agreement with the experimental data points.

In interpreting the light-matter interactions in our nanotube system, one might intuitively treat the nanoribbons as a collection of individual nanotubes that form a phased array. As the cut angle $\theta$ decreases, the nanotubes become longer and the nanoribbon resonance redshifts. This picture is not accurate, and a full understanding of our system requires a treatment of our nanotube system as a bulk anisotropic medium. To differentiate between the nanotube phased array and anisotropic medium pictures, we measure the peak transmission amplitude of nanoribbons as a function of the polarization angle $\psi$ of the incident free-space light, at the nanoribbon resonance frequency (Figure 5a). We consider moderately doped nanoribbons with a cut angle $\theta = 45°$ and $L = 95$ nm, though the observed effect is general to other geometric parameters (Figure S4). With the nanotube phased array concept, we would expect maximum electric field coupling to the nanoribbons to occur when the incident field is polarized along the nanotube axis (i.e., $\psi = 0°$). However, we instead measure maximum coupling when $\psi = 25°$ (Figure 5b). This angular deviation is also captured by full-wave simulations (Figure 5c).

To explain this coupling phenomenon, we treat the SWCNTs as a continuous anisotropic film and calculate, with the transfer matrix method, the electric fields of hyperbolic plasmons ($\vec{E}_{\text{HPM}}$) propagating 45° relative to the nanotube axis. The frequency is set as the nanoribbon resonance frequency, and the ellipsometry data for moderately-doped films are used as the dielectric tensor $\overleftrightarrow{\varepsilon}$. We then calculate the dielectric polarization of the nanotubes as $\vec{P} = \varepsilon_0(\overleftrightarrow{\varepsilon} - \overleftrightarrow{I}) \cdot \vec{E}_{\text{HPM}}$, where $\overleftrightarrow{I}$ is the identity matrix. The orientation of $\vec{P}$ is plotted as the solid green arrow in Figure 5c and matches the polarization angle for maximal nanoribbon light coupling. This analysis indicates that light coupling to and strong mode confinement in the nanoribbons are a collective effect that arise from the hyperbolic effective medium response of the aligned SWCNTs, and they cannot be explained by the optical responses of individual SWCNTs. This experiment, together with our measurement of the hyperbolic plasmon dispersion, validates the use of an effective medium model to describe the optical properties of SWCNT films and establishes them as a hyperbolic metamaterial.

In summary, we demonstrate that horizontally aligned SWCNT films are a model system for studying tunable hyperbolic plasmons. The unit cell size of this HMM is limited only by the cross-sectional area of individual SWCNTs, and the hyperbolic dispersion wavelength range can be dynamically tuned across the MIR using gas doping. Micro-FTIR spectroscopy of SWCNT nanoribbons demonstrates strongly subwavelength light localization as a result of in-plane hyperbolic dispersion, and full-wave simulations and transfer matrix analyses elucidate the light-matter interactions in these strongly anisotropic effective media.

Future work will need to address sources of loss in aligned SWCNT metamaterials. For our films, the imaginary parts of the dielectric constants over the hyperbolic spectral ranges are fairly high, indicating absorption in these ranges. This absorption can originate from multiple sources, including SWCNT alignment disorder, charges and trapping sites from the substrate, doping inhomogeneity, and free carrier scattering. The latter can be mediated by defects in the nanotubes, the nanotube ends, and carrier-phonon scattering.[48] As such, our SWCNT films are currently best suited for concepts that utilize optical absorption, such as those involving thermal emission[49] and photodetection.[50] Losses in SWCNT films could potentially be reduced by assembling films with longer nanotubes, and by more uniformly controlling the nanotube doping level. Scanning near-field optical microscopy of plasmons propagating along isolated SWCNTs has shown very low losses, indicating the potential for bulk SWCNT plasmonic systems to also have low losses.[51]

This work opens up new opportunities to study nanoscale thermal radiation control,[49, 52] hyperbolic phonon-plasmon coupling,[31] and tunable epsilon-near-zero phenomena[47] in carbon nanotube-based systems. The ability to scale these systems to deep subwavelength dimensions will enable exploration of fully three-dimensional SWCNT hyperbolic resonators supporting extreme mode volumes[4] and two-dimensional hyperbolic systems comprising a monolayer or sub-monolayer of nanotubes.

**Author Contributions**

JF and AF conceived the experiment. PH and AF performed the experimental measurements. JR and SY interpreted the experimental data and performed the theoretical calculations. PH prepared the SWCNT sample. SS performed the ellipsometry analysis. All authors wrote and edited the manuscript. JR and SY contributed equally to this work.

**Acknowledgements**

We thank B. Zhao and G. Papadakis for feedback on the manuscript, and E. Narimanov, G. Naik, P. Avouris, and J. Rosenberg for helpful discussions. This work was supported by the Air Force Office of Scientific Research (AFOSR) Multidisciplinary University Research Initiative (MURI) under Award FA9550-16-1-0031, the National Science Foundation (NSF) under award number 1608525, and the Packard Fellowship Foundation. J.A.R. was supported by the Department of Defense through the National Defense Science and Engineering Graduate Fellowship Program. P.H. was supported by the Postdoctoral Research Abroad Program of the Ministry of Science and Technology Taiwan (National Science Council 106-2917-I-564-012).

**Figures**

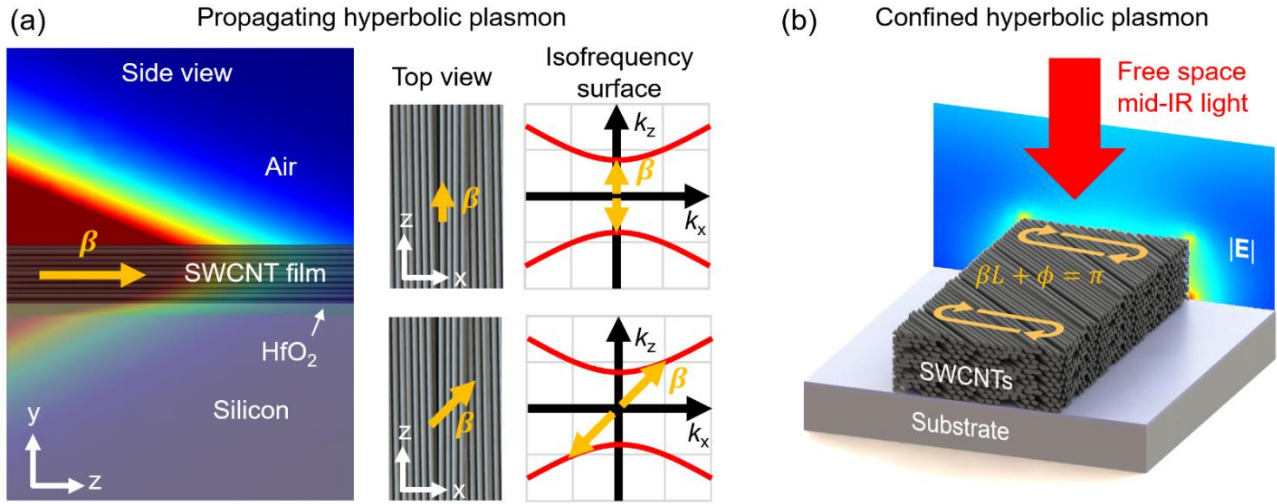

**Figure 1.** Schematics of hyperbolic plasmon modes in thin films and resonators. (a) Left: side view of the |**E**| field profile of the hyperbolic plasmon mode (HPM) propagating in a 95 nm-thick SWCNT film in air, on a hafnia and silicon substrate. Right: top views of in-plane HPM wavevectors propagating along and at an angle relative to the SWCNT alignment axes, and the corresponding isofrequency surfaces in wavevector space. When the HPM propagates at an angle to the SWCNT alignment axis, the in-plane propagation wavevector increases. (b) Schematic of the confined HPM in a nanoribbon resonator, which can be understood as a Fabry–Pérot cavity for propagating hyperbolic plasmons.

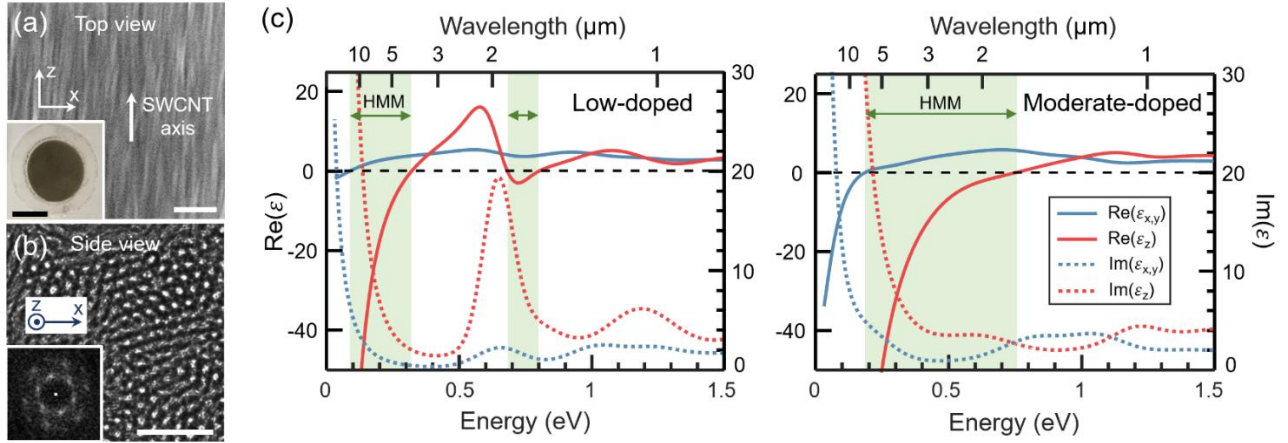

**Figure 2.** Ellipsometry measurements of aligned SWCNT films. (a) Top view SEM image of the aligned SWCNT film. Scale bar: 100 nm. Inset: photo of the aligned SWCNT film on the filter membrane. Scale bar: 1 cm. (b) Cross-sectional TEM image of the SWCNT film. Scale bar: 10 nm. Inset: Fourier transform of the TEM image. (c) Dielectric functions of 50 nm-thick SWCNT films, at a low doping level (left) and a moderate doping level (right). The substrate is silicon. The dielectric functions are extracted from the spectroscopic ellipsometry through the Mueller Matrix method (see Supporting Information). $\varepsilon_z$ is the complex dielectric function along the SWCNT axis, and $\varepsilon_{x,y}$ is the complex dielectric function along the plane perpendicular to the SWCNT axis. The light green boxes highlight the hyperbolic metamaterial (HMM) spectral range.

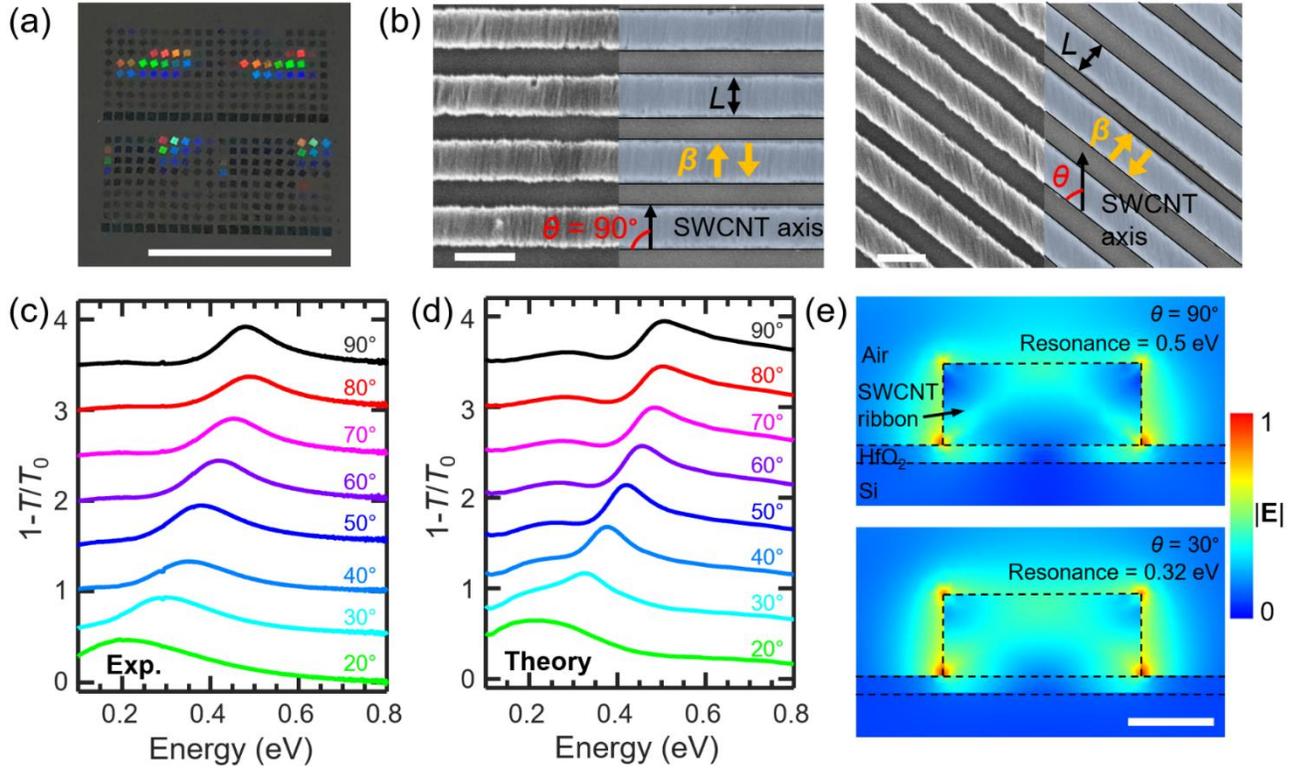

**Figure 3.** IR spectroscopy of SWCNT nanoribbons. (a) Patterned SWCNT film arrays on a HfO$_2$/Si substrate. Scale bar: 1 cm. (b) SEM images of SWCNT ribbon resonator arrays. Scale bars: 200 nm. The cut angle, $\theta$, is the relative angle between the SWCNT alignment axis and the edge of the etched nanoribbon. (c) Experimental and (d) theoretical IR transmission spectra from the micro-FTIR experiments on the moderate-doped patterned SWCNT films with $L = 230$ nm and different $\theta$ values. The theoretical spectra are obtained from FDTD simulations. The spectra are stack-plotted by adding vertical offsets of 0.5 between neighboring spectra. (e) Cross-sectional electrical field-strength maps of the SWCNT nanoribbons. The top nanoribbon is cut at $\theta = 90°$ and has a peak resonance at ~0.5 eV (2.5 μm) and the bottom nanoribbon is cut at $\theta = 30°$ and has a peak resonance at ~0.32 eV (3.9 μm). The dashed lines indicate material interfaces.

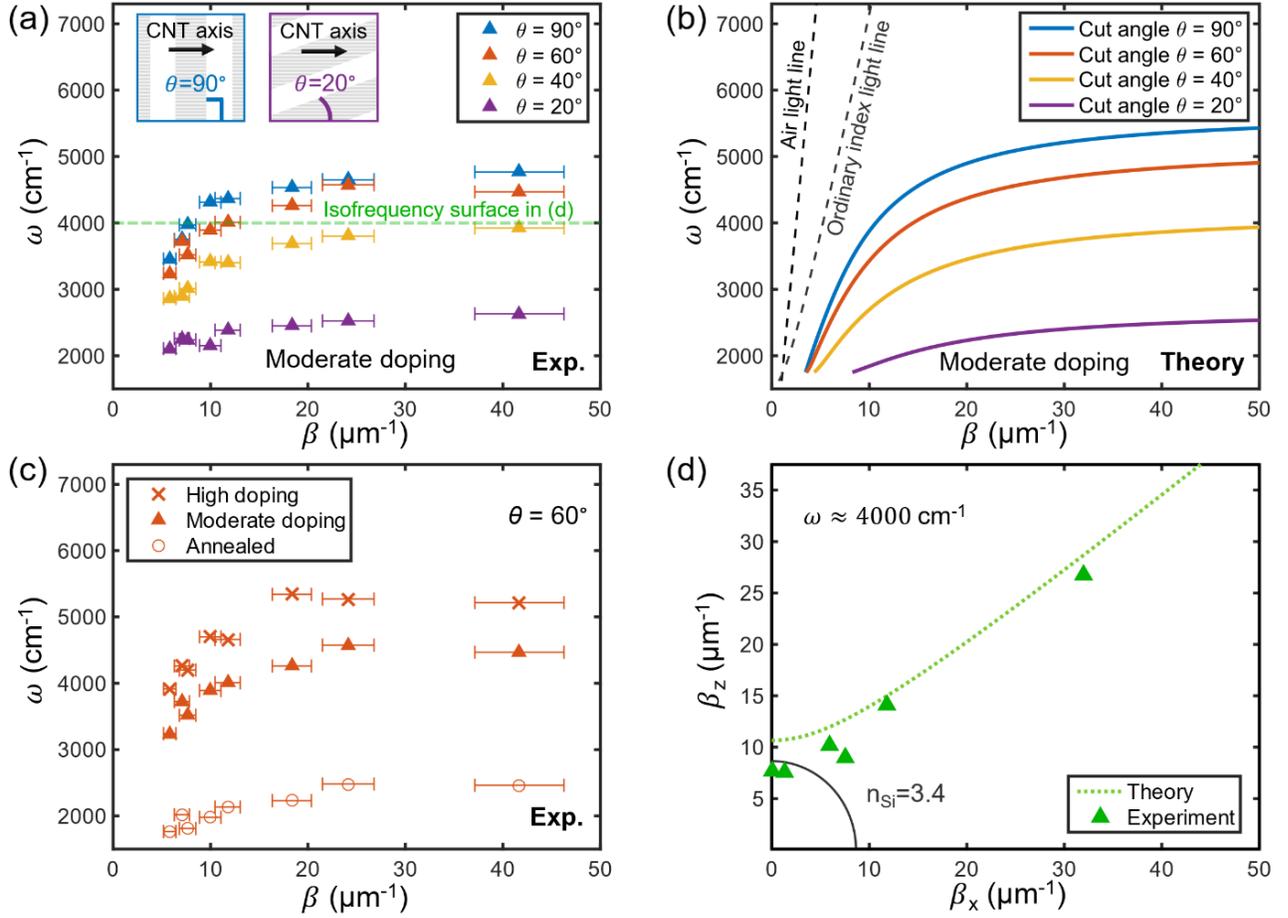

**Figure 4.** Dispersion of hyperbolic plasmon modes in SWCNT nanoribbons. (a) HPM dispersion plots for moderately-doped, 95 nm thick SWCNT nanoribbons, derived experimentally from the peak extinction frequency of HPM resonators that have a range of $L$ and $\theta$ parameters. The mode wavevector is defined to be $\beta = (\pi - \phi)/L$, where $\phi$ is determined numerically and has a level of uncertainty denoted by the error bars. (b) Theoretical dispersion relations of the HPM, calculated using transfer matrix theory and the ellipsometry dielectric functions from Figure 1c. (c) Dispersion relations of the HPM for $\theta = 60°$ and three doping levels, showing active tuning of the HPMs. (d) 2D in-plane isofrequency surface for moderately doped resonator arrays at $\omega = 4000 \text{cm}^{-1}$ (0.5 eV), using the data from (a). The theoretical curve (dashed) is calculated using transfer matrix theory and the ellipsometry data.

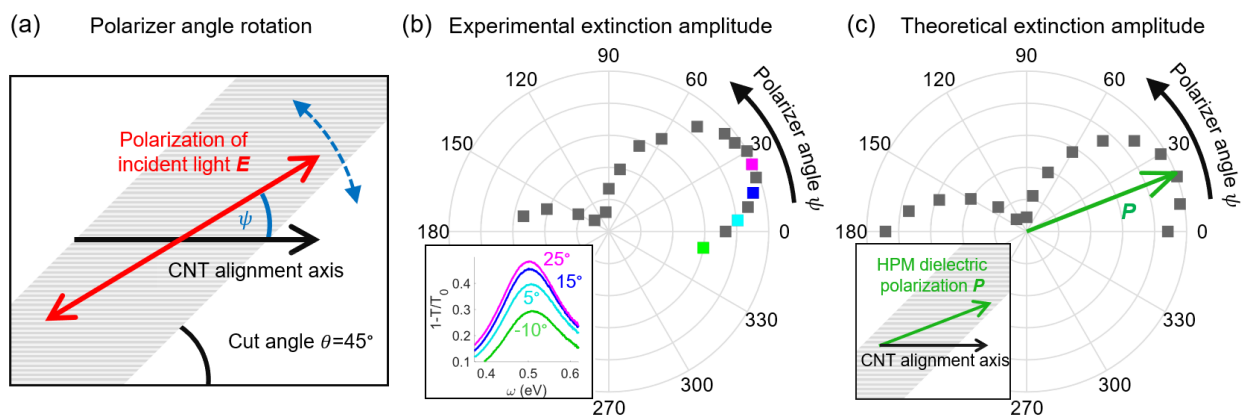

**Figure 5.** Dependence of extinction amplitude on incident wave polarization. (a) Schematic of a nanotube nanoribbon with varying incident light polarization. The peak extinction amplitude for a SWCNT nanoribbon array, cut at a 45° angle to the nanotube alignment axis, is recorded as a function of incident free-space light polarization. This incident light is linearly polarized with an angle $\psi$ relative to the SWCNT alignment axis. (b) Experimental extinction amplitude measurement of a 95 nm thick nanoribbon-resonator array with $L = 95$ nm, $\theta = 45°$ and a range of $\psi$. The polar plot is of peak extinction amplitude as a function of $\psi$ and shows a maximum peak extinction amplitude at $\psi = 25°$. Inset: extinction spectra for select $\psi$. (c) Theoretical peak extinction amplitude as a function of $\psi$, calculated using FDTD. The direction of the HPM dielectric polarization, **P**, is calculated using the transfer matrix method and coincides with the point of maximum peak extinction amplitude. Excellent agreement can be seen between theory and experiment.